\documentclass[12pt]{article}
\usepackage{amsmath,amssymb,amsthm,bm}
\usepackage{caption}
\usepackage{graphicx} 
\usepackage{float}

\numberwithin{equation}{section}

\include{PhysNote}

\begin{document}

\begin{flushright}
CERN-TH-2016-058\\

\bigskip

{\large To the memory of Raymond Stora}

\end{flushright}


\begin{center}

{\Large {\bf Relativistic causality and position space renormalization}
\footnote{\textit{Nuclear Physics} \textbf{B912} (2016) 79-87 (updated).}

\vspace{4mm}

 {\large Ivan Todorov} \\
{\small Theoretical Physics Department, CERN, CH-1211 Geneva 23, 
Switzerland} \\
{\small permanent address:} \\
\medskip
{\small Institute for Nuclear Research and Nuclear Energy \\ 
Bulgarian Academy of Sciences}
\\ {\small Tsarigradsko Chaussee 72, BG-1784 Sofia, Bulgaria} \\
\medskip
{\small ivbortodorov@gmail.com}}

\end{center}

\begin{abstract}

The paper gives a historical survey of the causal position space  
renormalization with a special attention to the role of Raymond Stora in 
the development of this subject. Renormalization is reduced to subtracting the
 pole term in analytically regularized primitively divergent Feynman amplitudes.
The identification of residues with ``quantum periods'' and their relation to 
recent developments in number theory are emphasized. We demonstrate the 
possibility of integration over internal vertices (that requires control over 
the infrared behavior) in the case of the massless $\varphi^4$ theory and 
display the dilation and the conformal anomaly.

\end{abstract}


\newpage

\tableofcontents

\vfill\eject

\section{Introduction}
As Raymond Stora had written\footnote{I thank Paul Sorba for providing me with
Stora's "bio0908" for the French Academy.}  in his inimitable ironic style, he
had \textit{contributed to the "useful physics"} (in his work with P. Moussa 
on angular distributions in 2-particle reactions) {\it as well as to the 
"useless" quantum field theory (QFT), including the analysis of analytic 
properties of scattering amplitudes which follow from the causality principle}
 -- in joint work with Bros, Epstein, Glaser, Messiah (see, e.g., [EGS]). 
Not surprisingly, our discussions at CERN were devoted to the ``useless'' part. 
 
 Perturbative ultraviolet renormalization in QFT was originally worked out for 
momentum space integrals beginning with a high energy cutoff. But a causal 
position space approach has also been developed concurrently by Ernst Stueckelberg, a Swiss student of Sommerfeld, starting in the early forties (after a 1938 paper in German, anticipating the abelian Higgs-Kibble model, he switched to French - see [S45, S46, SR, SP]). This was taken up by a (French reading) mathematician, N. N. Bogolubov [B], who set himself to master QFT (while mobilized to work -- with many others -- on the Russian atomic project).   The Russian work on renormalization (referred to in the book [BS] -- see, in particular, [St]), 
perfected by Hepp [He], Zimmermann and Lowenstein [Z, LZ] (resulting in the 
/incomplete/ acronym BPHZ) is still substantially using the traditional momentum space picture. Even Epstein and Glaser [EG], who set the stage for the position space renormalization program based on locality, were proving Lorentz invariance of time-ordered products working in momentum space. It was only in [PS] -- another famous unpublished preprint of Raymond's --  that the problem was translated into a cohomological position space argument (see the historical survey in [G-BL]). This led gradually to viewing renormalization as a problem of extending distributions defined originally for non-coinciding arguments, an approach that, in the words of Stora [S], "from a philosophical point of view, does not require the use -- and the removal -- of regularizations". The tortuous path 
from p- to x-space renormalization can be viewed, in modern parlance, as a 
duality transformation (the good old Fourier integral) mapping a large 
momentum onto a small distance problem. As relativistic causality does not 
require the existence of a Poincar\'e invariant vacuum state, the 
Stueckelberg-Bogolubov-Epstein-Glaser-Stora position space approach turned out
 to be the only one suited for the study of perturbative QFT on a curved 
background (which began flourishing during the last twenty years or so - see
\cite{HW, FR} for recent reviews and references). 

Our collaboration started with Raymond reading Sect. 3.2 of the first volume of
 H\"ormander's treatise [H] and pointing out 
that it is tailor-made for renormalization of a massless theory. It is based on
 the observation that a density like
\begin{equation}
\label{densG}
\textbf{G}_\ell(x) := G_\ell(x)\frac{d^4x}{\pi^2} = \frac{1}{x^{2\ell}}\frac{d^4x}{\pi^2}
\end{equation} 
is a meromorphic distribution valued function of $\ell$ with simple poles (at 
$2\ell=4, 5, 6, ...$ above).  Subtracting the pole term, say at $\ell=2$, we 
find a renormalized amplitude $G^R_2$ defined up to a distribution
 with support at the origin. The ambiguity can be restricted by demanding that 
this distribution has the same degree of homogeneity as the function $G_2$ away
 from the origin (in our case $-4$). The resulting $G^R_2$ is associate 
homogeneous of degree $-4$ and order one. More generally, a logarithmically 
divergent density $\textbf{G}$ of an N-dimensional argument $\vec{x}$ defines 
an \textit{associate homogeneous distribution $G$ of degree $-N$ and order $n$}
 if
\begin{equation}
\label{assocHom}
\lambda^N G(\lambda \vec{x})= G(\vec{x}) +\sum_{j=1}^n R_j(G)(\vec{x}) \frac{(\ln\lambda)^j}{j!}, \, \lambda > 0,
\end{equation}
where the distributions $R_j(G)$ can be viewed as generalized residues:
\begin{equation}
\label{generRes}
R_j(G)=Res[(\mathcal{E}+N)^{j-1}G(\vec{x})]\ ,\quad \mathcal{E}= \sum_{\alpha=1}^N x^\alpha\partial_\alpha,
\end{equation}
satisfying
\begin{equation}
\label{assocHomRes}
\lambda^N R_j(G)(\lambda\vec{x}) = R_j(G)(\vec{x}) + \sum_{i=j+1}^n R_i(G)(\vec{x}) \frac{(\ln\lambda)^{i-j}}{(i-j)!}, \, \lambda > 0. 
\end{equation}
For a Feynman amplitude corresponding to a connected graph with $V$ vertices $N=4(V-1)$. The order $n$ of associate homogeneity corresponds to the number of (sub)divergences of the amplitude. One proves that only the coefficient to the highest power of the logarithm,
\begin{equation}
\label{rn}
R_n(G) =res[(\mathcal{E}+N)^{n-1}G(\vec{x})]\delta(\vec{x})\ ,
\end{equation}
is independent of the ambiguity of renormalization.    

\medskip

\section{Causal factorization of extended Feynman amplitudes}
We start by sketching the recursive procedure of extending/renormalizing euclidean picture Feynman amplitudes based on causal factorization.  

Denote the propagator between the points $x_i$ and $x_j$ of $\mathbb{R}^4$ by 
$G_{ij}=G_{ij}(x_{ij}), \, x_{ij}= x_i-x_j$. We assume it to be a (bounded at infinity) smooth 
function away from the origin (i.e. off the diagonal $x_i=x_j$). In the case of a massless theory, treated in 
\cite{NST12, NST}, it is a rational homogeneous function of the type:
\begin{equation}
\label{prop}
G_{ij}(x)=\frac{P_{ij}(x)}{(x^2)^{m_{ij}}} \ , \quad  x^2 = \sum_{\alpha = 1}^4 (x^\alpha)^2, \, \, m_{ij}\in \mathbb{N},
\end{equation}
where $P_{ij}(x)$ is a homogeneous polynomial in the components $x^\alpha$ of $x$. (In a scalar QFT $P_{ij}=const, \,
m_{ij}=1$.) For the formulation of the \textit{principle of causal factorization} one does not need the special form of the propagator. It sets a condition on a recursive (with respect to the number of vertices) procedure of \textit{renormalization} (i.e. extension) of Feynman amplitudes. 
\smallskip

Let the index set $I = \{1,\ldots ,n\}$ of $\Gamma$ be split into any two non-empty non-intersecting subsets
$$
I = I_1 \cup I_2 \ (I_1 \ne \emptyset \, , \ I_2 \ne \emptyset) \, , \ I_1 \cap I_2 = \emptyset \, .
$$
Let $\mathcal{C}_{I_1,I_2}=\{(x_i)\in \mathbb{R}^{4n} \equiv (\mathbb{R}^4)^{\times n}; x_{j_1} \neq x_{j_2} \, \mbox{for} \, j_1 \in I_1, j_2\in I_2 \} (= 
\mathcal{C}_{I_2,I_1})$. Let further $G_1^R$ and $G_2^R$ be the renormalized 
distributions associated with the subgraphs whose vertices belong to the 
subsets $I_1$ and $I_2$, respectively. We demand that for each such splitting 
the extended \textit{euclidean} distribution $G_{\Gamma}^R$ exhibits the 
{\it factorization property}:
\begin{equation}
\label{G12R}
G_{\Gamma}^R = G_1^R \left(\prod_{i \in I_1 \atop j \in I_2} G_{ij} \right) G_2^R \quad \mbox{on} \quad \mathcal{C}_{I_1,I_2} \, ,
\end{equation}
where $G_{ij}$ are factors (propagators) in the Feynman amplitude $G_{\Gamma}$ which are smooth in $\mathcal{C}_{I_1,I_2}$ and can therefore be viewed as multipliers.

\textbf{Remark 1}. In the Lorentzian signature case one demands that the 
points indexed by the set $I_1$ precede those of $I_2$ and uses Wightman 
functions instead of $G_{ij}$ in the counterpart of (\ref{G12R}) (see Sect. 
2.2 of \cite{NST}).

\smallskip

In the case of a massless theory we add to this basic physical requirement two more {\it mathematical conventions} (MC) which restrict substantially the set of admissible renormalizations.

(MC1) {\it Renormalization maps rational homogeneous functions onto associate homogeneous distributions of the same degree of homogeneity;
it extends associate homogeneous distributions defined off the small diagonal to associate homogeneous distributions of the same degree}
(but possibly of higher order) {\it defined everywhere on ${\mathbb R}^N$.}

(MC2)  {\it The renormalization map commutes with multiplication by polynomials.} If we extend the class of our distributions by allowing
multiplication with smooth functions of no more than polynomial growth (in the domain of definition of the corresponding functionals),
then this requirement will imply commutativity of the renormalization map with such multipliers.


\smallskip

The induction is based on the following \textit{diagonal lemma}.

{\bf Proposition 1}. {\it The complement $C(\Delta_n)$ of the small diagonal is the union of all $\mathcal{C}_{I_1,I_2}$ for all pairs of disjoint $I_1,I_2$ with $I_1 \cup I_2 =\{1,\dots,n\}$, i.e.,
$$
C(\Delta_n) \, = \,
\mathop{\bigcup}\limits_{I_1 \dot \cup I_2 \, = \, \{1,\dots,n\}} \mathcal{C}_{I_1,I_2} \,.
$$}

\bigskip

\noindent {\bf Proof.} Let $(x_1 , \ldots , x_n) \in C(\Delta_n)$. Then there are at least two different points $x_{i_1} \ne x_{j_1}$. We define $I_1$ as the 
set of all indices $i$ of $I=\{1,\dots,n\}$ for which $x_i \ne x_{j_1}$ and 
$I_2 := I \backslash I_1$. Hence, $C(\Delta_n)$ is included in the union of all
 such pairs. Each $\mathcal{C}_{I_1,I_2}$, on the other hand, is defined to 
belong to $C (\Delta_n)$. This completes the proof of our statement.

\textbf{Remark 2}. For a more general combinatorial "diagonal lemma" that 
serves both the euclidean and the Minkowski space framework allowing to 
complete each step of the renormalization by the extension of a distribution 
defined outside the full diagonal - see Theorem A1 of [NST].  
\medskip                        
\section{Renormalization of primitively divergent \\
amplitudes}
The above recursive procedure allows to reduce the elimination of divergences to the renormalization of primitively divergent graphs. We shall again survey this step in the case of a euclidean massless QFT. A Feynman amplitude $G(\vec{x})$ is then a homogeneous function of $\vec{x}\in \mathbb{R}^N$. It is {\it superficially divergent} if $G$ defines a density in $\mathbb{R}^N$ of a non-positive degree of homogeneity:
\begin{equation}
\label{homdens}
G(\lambda \vec{x})\, d^N \lambda x = \lambda^{-\kappa} G(\vec{x})\, d^N x\ ,\quad \kappa\geq 0\   \quad  (\lambda > 0)\,;
\end{equation}
$\kappa$ is called the (superficial) {\it degree of divergence}. 

{\bf Proposition 2.} {\it For any primitively divergent $G(\vec{x})$ and smooth (semi)norm $\rho(\vec{x})$ on $\mathbb{R}^N$ (allowed to vanish on a cone of lower dimension) one has
\begin{equation}
\label{Res} [\rho(\vec{x})]^\epsilon G(\vec{x})
-\frac{1}{\epsilon}(Res\, G)(\vec{x}) = G^\rho(\vec{x}) +
O(\epsilon).
\end{equation}
Here $Res\,G$ is a distribution with support at the origin. Its calculation is reduced to the case $\kappa = 0$ of a logarithmically divergent graph by using the identity
\begin{equation}
(Res\,G)(\vec{x})=\frac{(-1)^\kappa}{\kappa!}\,\partial_{i_1}...\partial_{i_\kappa} Res\,(x^{i_1}...x^{i_\kappa} G)(\vec{x})
\end{equation}
where summation is assumed (from 1 to N) over the repeated indices $i_1, ..., i_\kappa$. If $G$ is homogeneous of degree $-N$ then
\begin{equation}
\label{res}
(Res\, G)(\vec{x}) = res\, (G)\, \delta(\vec{x}) \quad (\mbox{\rm for} \ \partial_i(x^iG)=0)\, .
\end{equation}
Here the numerical residue $res\, G$ is given by an integral over the hypersurface $\Sigma_\rho =\{\vec{x}| \,\rho(\vec{x}) = 1\}$:
\begin{equation}
res\, G =\frac{1}{\pi^{N/2}}\int_{\Sigma_\rho} G(\vec{x})\sum_{i=1}^N(-1)^{i-1} x^i dx^1\wedge ... \hat{dx^i}...\wedge dx^N, \,
\end{equation}
(a hat over an argument meaning, as usual, that this argument is omitted).
The residue $res\, G$ is independent of the (transverse to the dilation) surface $\Sigma_\rho$ since the form
in the integrand  is closed in the projective space $\mathbb{P}^{N-1}$.}

We note that N is even, in fact divisible by 4, so that $\mathbb{P}^{N-1}$ is orientable.

\textbf{Remark 3}. The use of a homogeneous (semi)norm as a regulator (a 
relative of {\it analytic regularization} \cite{Sp}) is more flexible than 
dimensional regularization and should be also applicable in the presence of a 
chiral anomaly.

The functional $res\, G$ is a {\it period} according to the definition of 
Kontsevich and Zagier \cite{KZ}. The convention of accompanying the 4D volume 
$d^4x$ by a $\pi^{-2}$ factor ($2\pi^2$ being the volume of the unit sphere 
$\mathbb{S}^3$ in four dimensions) helps display the number theoretic character
 of residues. For one and two-loop graphs in a massless theory they are just 
rational numbers. For three, four and five loops in the $\varphi^4$ theory all 
residues are integer multiples of $\zeta(3), \zeta(5)$ and $\zeta(7)$, 
respectively. The first double zeta value, $\zeta(3, 5)$, appears at six loops 
(with a rational coefficient) (see the census of Schnetz who calls such residues \textit{quantum periods} \cite{Sch}). All {\it known} residues were (up to 2013) rational linear combinations of multiple zeta values of overall weight not exceeding $2\ell-3$ \cite{BK, Sch}. A seven loop graph was recently demonstrated \cite{P, B14} to involve {\it multiple Deligne values} -- i.e., values of {\it hyperlogarithms} at sixth roots of unity. An infinite series of $\ell$-loop primitive $\varphi^4$ 4-point \textit{zig-zag graphs} were conjectured by Broadhurst and Kreimer [BK] and proven by Brown and Schnetz [BS12] to be proportional to   
$\zeta(2\ell-3)$ with calculable rational coefficients (equal to ${2\ell-2\choose \ell-1}$ \, for \, $\ell=3, 4$ -- see \cite{T} for an elementary derivation and further references).

\medskip

\section{Integration over internal vertices. \\ 
Completed $\varphi^4$ vacuum graphs}
In the adiabatic procedure of Bogolubov et al. all vertices are treated as external: each coupling constant $g$ is substituted by a vanishing at infinity test function $g(x)$. This is essential for the formulation of causal factorization. Integration over internal vertices corresponds to the adiabatic limit ($g(x)\rightarrow g\neq 0$) and does not keep track of localization. It is rewarding to understand that such an integration commutes with renormalization and hence does not pose a problem in a conformally invariant theory like $\varphi^4$ in $D=4$, \cite{GGV, T15} (thus elucidating an old result, \cite{LZ}).

We shall sketch the basic idea using Schnetz's \textit{vacuum completion} $\bar{\Gamma}$ of a 4-point graph $\Gamma$
(in which the four external edges are joined together in a new "vertex at infinity" - \cite{Sch, S14}). The introduction of this concept is justified by the following result (Proposition 2.6 and Theorem 2.7 of \cite{Sch}):

{\bf Proposition 3.} {\it A 4-regular vacuum graph $\bar{\Gamma}$ (with five or more vertices) is said to be completed primitive if the only way to split it by a four edge cut is by splitting off one vertex. A 4-point Feynman amplitude corresponding to a connected 4-regular graph $\Gamma$ is primitively divergent iff its completion $\bar{\Gamma}$ is completed primitive. All 4-point graphs with the same primitive completion have the same residue.}
  
There are infinitely many primitive 4-point graphs (while there is a single 
primitive 2-point self-energy graph).

{\bf Proposition 4}. {\it The period of a completed primitive graph $\bar{\Gamma}$ is equal to the residue of each 4-point graph $\Gamma = \bar{\Gamma} - v$ (obtained from $\bar{\Gamma}$ by cutting off an arbitrary vertex $v$). The resulting common period can be evaluated from $\bar{\Gamma}$ by choosing arbitrarily three vertices $\{0, e \, (s.t. \, e^2=1), \infty\}$, setting all propagators corresponding to edges of the type $(x_i, \infty)$ equal to 1 and integrating over the remaining $n-2$ vertices of $\Gamma$ ($n=V(\Gamma)$):
\begin{equation}
\label{Per}
Per(\bar{\Gamma})\equiv res(\Gamma)=
\int \Gamma(e, x_2, ..., x_{n-1}, 0)\prod_{i=2}^{n-1} \frac{d^4x_i}{\pi^2}\ .
\end{equation}}
{\it Sketch of proof}. For a given choice of the vertex at infinity (\ref{Per}) follows from (\ref{res}). The independence of the choice of the point at 
infinity follows from conformal invariance. We note, for instance, that 
the conformal inversion $I_r: x_i\rightarrow \frac{x_i}{x_i^2}, i=2, ..., n$, 
exchanges the (arbitrarily chosen) $x_n=0$ and $\infty$ while the integral 
remains invariant since
\begin{equation}
\label{inversion}
I_r: \frac{1}{x_{ij}^2}\rightarrow \frac{x_i^2 x_j^2}{x_{ij}^2}\ ,\quad d^4x\rightarrow \frac{d^4x}{(x^2)^4}\ .
\end{equation}

It is the freedom of choice of the vertices to which one ascribes the values $0, e, \infty$ in Propositon 4 (as a consequence of conformal invariance) that guarantees the commutativity between renormalization and integration with respect to internal vertices. One can illustrate this fact on the four-loop graph of Fig. 1 with a single internal vertex $x$ (the black dot in the middle of the figure). 
\begin{figure}[htb]
\centering
\includegraphics[width=0.7\textwidth]{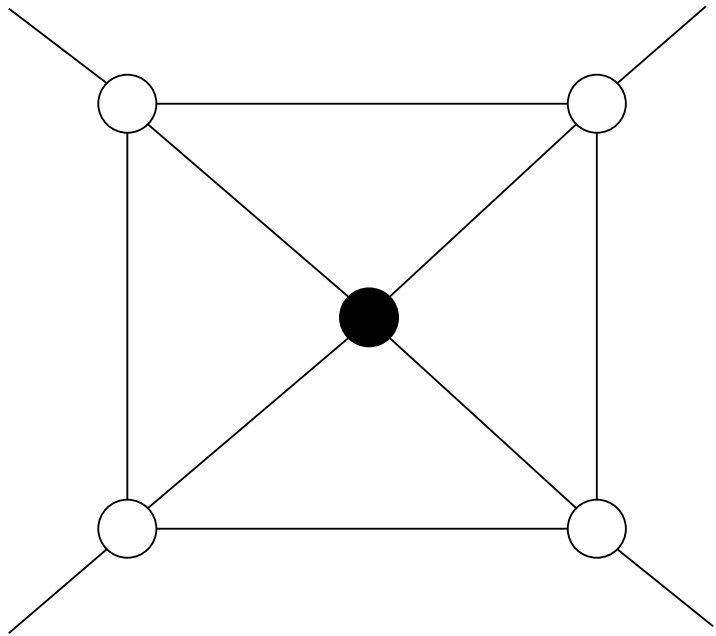}
\caption{\footnotesize{Four-loop 4-point graph $G_4$}}
\label{Fig1}
\end{figure}
The simplest way to calculate the residue of the corresponding amplitude $G_4$ 
consists in setting $x=0$ (rather than integrating in x). The result appears as
 a special case (for $\ell = 4$) of the \textit{wheel with $\ell$ spokes} 
expressed in terms of the classical polylogarithm \cite{S14, T}:
\begin{equation}
\label{lspokes}
resG_\ell = {2\ell-2\choose \ell-1} Li_{2\ell-3}(1) = {2\ell-2\choose \ell-1} 
\zeta(2\ell-3) \, (resG_4=20\zeta(5)).
\end{equation} 
If, on the other hand, one first integrates with respect to $x$ (expressing $G_4$ in terms of the Bloch-Wigner dilogarithm) then the residue is calculated in terms of multipolylogarithms of higher depth \cite{T} but the final answer is the same
- as a consequence of conformal invariance.

\section{Dilation and conformal anomalies}

The renormalized Feynman amplitude $G(x_1, ..., x_4)$ of an arbitrary 
primitively divergent 4-point graph is an associate homogeneous distribution of
 order one (and degree twelve in the generic case when there is a single 
external edge at each external vertex):
\begin{equation}
\label{4ptdil}
\lambda^{12}\, G(\lambda x_1, ..., \lambda x_4) = G(x_1, ..., x_4) + res(G)\, \delta(x_{12}) \delta(x_{23}) \delta(x_{34}) f(\lambda), \, \,
\end{equation} 
where $f$ is a 1-cocycle (normalized by $f^\prime(1)=1$):
\begin{equation}
\label{1cocycle}
f(\lambda_1 \lambda_2) = f(\lambda_1) + f(\lambda_2) \Rightarrow f(\lambda) = \ln\lambda.
\end{equation}

Graphs with subdivergences give rise to associate homogeneous amplitudes of higher order. The generalized residue $R_n(G)$ (\ref{rn}) appearing as coefficient to the highest power of $\ln\lambda$ can be computed in terms of the residues 
of the divergent subgraphs and of the corresponding quotient graphs. We shall illustrate this fact on the example of the graph on Fig. 1 in which the central point is substituted by a generic primitively divergent 4-point subgraph with amplitude $S(y_1, ..., y_4)$
\begin{equation}
\label{GS}
G_S(x_1, ..., x_4)=\int S(y_1, ...,y_4)\prod_{i=1}^4 \frac{d^4y_i}{\pi^2 (x_i-y_i)^2}\ .
\end{equation}
The dilation law for $S$,
\begin{equation}
\label{dilS} \lambda^{12}S(\lambda\vec{y})=S(\vec{y}) + res(S)\,\delta(\vec{y})\ln\lambda
\end{equation}
implies that the dilation anomaly of $G_S$ for non-coinciding arguments is
\begin{equation}
\label{anomGS}
\lambda^{12}G_S(\lambda x_1, ..., \lambda x_4)-G_S(x_1, ..., x_4)= G_4(x_1, ..., x_4)\, res(S)\ln\lambda\ ,
\end{equation}
where $G_4$ is given by 
\begin{equation}
\label{G4}
G_4(x_1, ..., x_4) = \frac{1}{x_{12}^2 x_{23}^2 x_{34}^2 x_{14}^2}\int\prod_{i=1}^4\frac{1}{(x_i-x)^2} \frac{d^4x}{\pi^2}.
\end{equation}
It follows that the coefficient $res_2(G_S)$ to $(\ln\lambda)^2$, which is 
independent of the renormalization ambiguity, is given by the product of 
residues:
\begin{equation}
\label{resGresS}
res_2(G_S)= res(G_4)\, res(S) \qquad (res(G_4) = 20\, \zeta(5))\ .
\end{equation}

A renormalized primitively divergent 4-point graph also has a calculable 
conformal anomaly. Under the special conformal transformation
\begin{equation}
\label{gc}
g_c x = \frac{x+c x^2}{\omega(c, x)}\ ,\quad (dg_c x)^2=\frac{dx^2}{\omega(c, x)^2}\ ,\quad
\omega(c, x)= 1 + 2cx + c^2 x^2.
\end{equation}
the renormalized amplitude $G$ obeys the following counterpart of (\ref{4ptdil}):
\begin{eqnarray}
\label{conf}
\frac{G(g_cx_1, ..., g_cx_4)}{\prod_{i=1}^4\omega^3(c, x_i)}= G(x_1, ..., x_4) - res(G)\, \delta(x_{12}) \delta(x_{23}) \delta(x_{34})\ln\omega(c, x_j), \, \nonumber \\ 
j\in(1, 2, 3, 4). 
\end{eqnarray} 
The $\delta$-function ensures that the result is independent of the choice of $j$ in the last factor. The
cocycle condition that implements the group law is satisfied because of the identity
\begin{equation}
\label{cocycle}
\omega(c_1+c_2, x) = \omega(c_1, x) \omega(c_2, g_{c_1}x)\ .
\end{equation}

\smallskip

\section{Outlook} 
There is a parallel between studying renormalzation of a \textit{massless} QFT and neglecting friction by the founders of modern physics  -- starting with Galileo. Both idealizations allow to grasp the essence of the problem. Introducing friction in classical mechanics, and masses in the analysis of small distance behavior seems to be just adding technical details to the general picture. Raymond, however, \textit{did worry} about masses in QFT renormalization. Recent work \cite{ABW, BKV} on a simple 2-point amplitude with arbitrary non-zero masses illustrates the arising complications. Nevertheless, the causal
 position space approach to renormalization wworks in this general case as well \cite{N, VG}. 
  
\smallskip

The study of Feynman periods, an essential ingredient of renormalization 
theory (Sect. 3), is bringing a new insight in a lively area of number theory 
(see \cite{B15, PS16} for recent developments in this subject).

\medskip

As we see, and work in the last couple of decades, surveyed, e.g. in  
\cite{D, T16}, amply confirms, "useless" local QFT continues to serve both
 high energy physics and its healthy interaction with modern mathematics.

\bigskip

I thank the Theoretical Physics Department of CERN for hospitality in February-March 2016 when this paper has been essentially completed. The author's work has been supported in part by Grant DFNI T02/6 of the Bulgarian National Science Foundation.

\end{document}